\documentclass{ws-ijmpa}

\begin{document}

\markboth{S. Atashbar Tehrani, Ali N. Khorramian and A. Mirjalili}
{Nuclear Parton Densities and Structure Functions}

%%%%%%%%%%%%%%%%%%%%% Publisher's Area please ignore %%%%%%%%%%%%%%%
%
\catchline{}{}{}{}{}
%
%%%%%%%%%%%%%%%%%%%%%%%%%%%%%%%%%%%%%%%%%%%%%%%%%%%%%%%%%%%%%%%%%%%%

\title{NUCLEAR PARTON DENSITIES AND STRUCTURE FUNCTIONS\\}

\author{\footnotesize S. Atashbar Tehrani\footnote{
atashbar@ipm.ir}}

\address{
 Physics Department, Persian Gulf University, Boushehr,
Iran\\
Institute for Studies in Theoretical Physics and Mathematics
(IPM)\\P.O.Box 19395-5531, Tehran, Iran}

\author{Ali N.Khorramian}

\address{Physics Department, Semnan University, Semnan, Iran\\
Institute for Studies in Theoretical Physics and Mathematics
(IPM)\\P.O.Box 19395-5531, Tehran, Iran}

\author{A. Mirjalili}

\address{Physics Department, Yazd University, Yazd, Iran\\
Institute for Studies in Theoretical Physics and Mathematics
(IPM)\\P.O.Box 19395-5531, Tehran, Iran}

\maketitle

\pub{Received (Day Month Year)}{Revised (Day Month Year)}

\begin{abstract}
We calculate nuclear parton distribution functions (PDFs), using
the constituent quark model. We find the bounded valon
distributions in a nuclear to be related to free valon
distributions in a nucleon. By using improved bounded valon
distributions for a nuclear with atomic number $A$ and the
partonic structure functions inside the valon, we can calculate
the nuclear structure function in $x$ space. The results for
nuclear structure-function ratio $F_2^A/F_2^D$ at some values of
$A$ are in good agreement with the experimental data.

\keywords{Constituent quark model; Bounded valon; EMC.}
\end{abstract}

\section{INTRODUCTION}
The difference of the structure function for a bounded nucleon
which measured on a nuclear target, first was found by the
European Muon Collaboration (EMC). It then compared with a
quasi-free nucleon measured on deuterium target. This phenomenon
is called the EMC effect. Due to this effect, much experimental
and theoretical efforts have been done  to investigate  the
features of the bounded nucleons in a nucleus.\\

We need a model to account for the structure of a free nucleon and
will use it as a base for studying its bounded features. Such a
model should first include the relationship between the
constituent quark and the current quark. It should also be in a
good agreement with the DIS data on a free nucleon. Some papers
can  be found
 in which the valon model has been used to extract new
information for  parton distributions and  hadron structure
functions \cite{some}. In what follows we employ the expression of
the nucleon structure function for a free nucleon used by Hwa and
Zahir\cite{Hwa}
\begin{eqnarray}
F_N(x,Q^2)=\sum_c\int_x^1 dy\; G^{free}_c(y)F_{c/N}(x/y,Q^2)\;,
\end{eqnarray}
here the summation runs over all constituent quarks or valons and
$G^{free}_c(y)$ is the free valon distributions in the nucleon
which has nothing to do with a probe. In Eq.(1) $F_N(x,Q^2)$ and
$F_{c/N}(x/y,Q^2)$ are respectively the nucleon and valon
structure functions which are depends on $Q^2$ and the nature of
the probe. The distribution function $G^{free}_c(y)$ governed by
nonperturbative QCD, i.e., the implication of the factorization
assumption. In considering the constituent quark model in nuclear
structure function, there are two kind of valons: bounded valons
in nuclei and free ones in nucleon. There is a relation between
these two distributions which will be discussed in more details in
next section.

\section{PROPERTIES OF CONSTITUENT QUARKS IN A FREE AND BOUNDED NUCLEON}

A free nucleon in the constituent-quark model can be considered as
a composite system consisting of three constituent quarks. Their
distribution in a nucleon satisfies normalized condition
\begin{equation}
\int_0^1G_c^{free}(y)dy=1\;,
\end{equation}
and the momentum sum rule
\begin{equation}
\int_0^1G_c^{free}(y)ydy=\frac{1}{3}\;.
\end{equation}

The sum rule means that each constituent quark carries, on an
average, one-third of the longitudinal momentum of a nucleon. The
form of the distribution $G^{free}_c$ is determined by the
interactions among the constituent quarks and is independent of
probe. For bounded nucleon we suppose that it still consists of
three constituent quarks in the same manner as a free nucleon,
i.e., the normalized condition is,
\begin{equation}
\int^1_0G_c^{bound}(y)dy=1\;.
\end{equation}

In addition, not only  the binding interactions with a few MeV do
not change the longitudinal momentum carried by each bounded
nucleon in DIS, but also they do not change the average
longitudinal momentum carried by each constituent quark.
Therefore, under the condition of the approximation it seems to be
reasonable to suppose that the momentum sum rule
\begin{equation}
\int_0^1G_c^{bound}(y)ydy=\frac{1}{3}\;,
\end{equation}
is valid in the same manner as a free nucleon. The distribution of
the $U$-constituent quark in a bounded nucleon which
simultaneously satisfies the conditions (4) and (5) can be written
as
\begin{eqnarray}
G_{U}^{bound }&=&\frac{y^{a+\alpha_U}(1-y)^{a+b+\alpha_U+\beta_U+1}}{%
Beta(a+\alpha_U+1,a+b+\alpha_U+\beta_U+2)}\;,
\end{eqnarray}
where $\alpha_{U}$ and $\beta_{U}$ are called distortion factor
for $U$-valon\cite{Zhu}. We can employ the same strategy for other
constituent quarks. If we supposed $\alpha_{U}=\beta_{U}=0$ then
we can find $G_{U}^{bound}=G_{U}^{free}$. Now we can introduce the
nuclear valon distribution in $A$-nuclei. Our ansatz for the
relation of valon distribution in $A$ nucleon and bounded valon
distribution is
\begin{equation}
G_{i/A}=w_{i}(y,A,Z)G_{i}^{bound }(y)\;,
\end{equation}
where $G_{i/A}$ is the $i$-valon distribution  in the $A$ nucleus.
We call $w_{i}(y,A,Z)$ as a weight function, which takes into
account the nuclear modification. The weight function for instance
of $U$-valon, $w_U$, is
\begin{eqnarray}
w_{_{U}}&=&1+(1-\frac{1}{A^{1/3}}%
)\frac{(u_{0}+u_{1}y+u_{2}y^{2}+u_{3}y^{3})}{(1-y)^{u_{4}}}\;.
\end{eqnarray}
We can have a similar weight functions for other constituent
quarks in nuclear. We can get eventually the nuclear valon
distribution  in the same method\cite{Hirai} as in following
\begin{eqnarray}
{\cal {G}}_{U}^{A}&=&\frac{ZG_{U/A}+NG_{D/A}}{A}\;,\\ {\cal
{G}}_{D}^{A}&=&\frac{ZG_{D/A}+NG_{U/A}}{A}\;,
\end{eqnarray}
and
\begin{equation}
{\cal {G}}_\Sigma ^{A}=w_{_{\Sigma} }G_{\Sigma /A}\;.
\end{equation}
Using the convolution integral we can obtain all of the nuclear
quark distribution functions,
\begin{eqnarray}
u_{v}^{A}(x,Q^{2})&=&2\int_{x}^{1}{\cal {G}}_{U}^{A}(y)f_{v}(z,Q^{2})\frac{dy}{y}\;,\\
d_{v}^{A}(x,Q^{2})&=&\int_{x}^{1}{\cal {G}}_{D}^{A}(y)f_{v}(z,Q^{2})\frac{dy}{y}\;,\\
\Sigma^{A}(x,Q^{2})&=&\int_{x}^{1}{\cal
{G}}_\Sigma^{A}(y)f_{\Sigma}(z,Q^{2})\frac{dy}{y}\;.
\end{eqnarray}

Now using Eqs.(12-14) we can extract nuclear sea quark
distribution. By substituting the obtained nuclear quark
distribution in the definition of nuclear structure function
as\cite{Hirai}
\begin{equation}
F_{2}^{A}=\frac{x}{9}[4u_{v}^{A}(x,Q^{2})+d_{v}^{A}(x,Q^{2})+12\overline{q}%
^{A}(x,Q^{2})]\;,
\end{equation}
we find that there are some unknown parameters which initially
appeared in Eq.(7). All of the unknown parameters can be obtained
by fitting the ration of $F_2^A/F_2^D$  to experimental
data\cite{E139-NMC}. Using the CERN subroutine
MINUIT\cite{minuit}, we can define a global $\chi^2$ for all the
experimental data points and find an acceptable fit.  In Fig.1 we
presented the analytical results for ratio of $F_2^{He}/F_2^D$ at
$Q^2=5\;GeV^2$ and compared it with available experimental data.

\begin{figure}
\centerline{\psfig{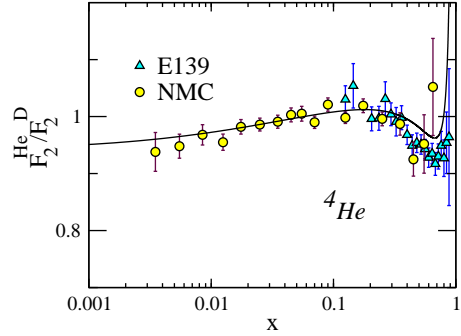}} \vspace*{8pt}
\caption{Ratio of $F_2^{He}/F_2^D$ at $Q^2=5\;GeV^2$, where the
data are taken at various $Q^2$ point.}
\end{figure}
\section*{Acknowledgments}
We acknowledge Institute for Studies in Theoretical physics and
Mathematics (IPM) to support financially this project. S.A.T
thanks Persian Gulf University for partial financial support to do
this project.

\end{document}